\documentclass[aps,prb,twocolumn,superscriptaddress,showpacs]{revtex4-1}

\usepackage{graphicx}
\usepackage{dcolumn}
\usepackage[usenames]{color}
\usepackage{bm}
\usepackage{units}
\usepackage[normalem]{ulem}
\usepackage{soul}
\setulcolor{red}

\definecolor{ultramarine}{rgb}{0.17, 0.34, 0.56}

\begin{document}

\title{Unveiling the stacking-dependent electronic properties of 2D ultrathin rare-earth metalloxenes family Ln$X_2$ (Ln~=~Eu, Gd, Dy; $X$ = Ge, Si) }

\author{Polina M. Sheverdyaeva}
%\email[]{polina.sheverdyaeva@trieste.ism.cnr.it}
\affiliation{Istituto di Struttura della Materia, Consiglio Nazionale delle Ricerche, I-34149 Trieste, Italy}

\author{Alexey~N.~Mihalyuk}
%\email[]{mih-alexey@yandex.ru}
%\cortext[cor2]{Corresponding author}
\affiliation{Institute of High Technologies and Advanced Materials, Far Eastern Federal University, 690950 Vladivostok, Russia}
\affiliation{Institute of Automation and Control Processes FEB RAS, 690041 Vladivostok, Russia}

\author{Jyh-Pin~Chou}
%\email[]{jpchou@cc.ncue.edu.tw}
\affiliation{Department of Physics, National Changhua University of Education No.1, Jinde Rd., Changhua City, Changhua County 50007, Taiwan}

\author{Andrey V. Matetskiy}
\affiliation{Istituto di Struttura della Materia, Consiglio Nazionale delle Ricerche, I-34149 Trieste, Italy}

\author{Sergey~V.~Eremeev}
\affiliation{Institute of Strength Physics and Materials Science SB RAS, Tomsk 634055, Russia}

\author{Andrey~V.~Zotov}
\affiliation{Institute of Automation and Control Processes FEB RAS, 690041 Vladivostok, Russia}

\author{Alexander~A.~Saranin}
\affiliation{Institute of Automation and Control Processes FEB RAS, 690041 Vladivostok, Russia}

\begin{abstract}
The studies of electronic effects in reduced dimensionality have become a frontier in nanoscience due to exotic and highly tunable character of quantum phenomena.
Recently, a new class of 2D ultrathin Ln$X_2$ metalloxenes composed of a triangular lattice of lanthanide ions (Ln) coupled with 2D-Xenes of silicene or germanene ($X_2$) was introduced and studied with a particular focus on magnetic and transport properties. However, the electronic properties of metalloxenes and their effective functionalization remain mainly unexplored.
Here, using a number of experimental and theoretical techniques, we trace the evolution of electronic properties and magnetic ground state of metalloxenes triggered by external perturbations.
We demonstrate that the band structure of Ln$X_2$ films can be uniquely modified by controlling the Xenes stacking, thickness, varying the rare-earth and host elements, and applying an external electric field. Our findings suggest new pathways to manipulate the electronic properties of 2D rare-earth magnets that can be adjusted for spintronics applications.
\end{abstract}

\maketitle
\clearpage

\section{Introduction}
The family of low-dimensional materials that hold 2D intrinsic magnetism has been impressively growing in recent years, expanding the scope of possible phenomena to be explored in 2D and enabling the development of novel devices \cite{ACS-Nano-Magnetic-Genome-2022,Nature-MAg-2D-vdW-2018,Nature-Nanotechnology-Mag-2D-2019}. Like graphene and transition-metal dichalcogenides revolutionized materials engineering, the discovery of 2D atomic magnets opened up enormous opportunities for applications and fundamental research.
The variety of 2D magnetic materials encompasses systems with various magnetic and electronic properties ranging from metals \cite{Nature-Fe3GeTe2-2018} to insulators \cite{Nature-2D-mag-2018}, and holds various magnetic orders, such as ferromagnetic \cite{Nature-Fe3GeTe2-2018}, antiferromagnetic \cite{Nature-Layer-Hall-2D-mag-2021}, and competing one \cite{Nano-Research-Competing-mag-Storchak-2020}. Structurally among 2D magnets there are van der Waals (vdW) crystals \cite{Nat-Nanotech-gating-CrI3-2018,Nat-Mat-mag-CrI3-2019,JPCM-2D-vdW-mag-2016} demonstrating transition temperatures at or above room temperature \cite{Nat-Nanotech-VSe2-vdW-2018,Nano-Letters-vdW-mag-2018}, unconventional superconductivity and flat-band physics \cite{Nature-flat-band-mag-2017}; single atomic layers having Ising-type magnetic order \cite{Nano-Letters-FePS3-2016}, layer-dependent ferromagnetism \cite{Nat-Nanotech-gating-CrI3-2018,Nature-intrinsic-FM-vdW-2017}; atomically thin lanthanide  metalloxenes \cite{Materials-Today-GdSi2-2019,Sokolov2020MH,JMCC_GdSGe2-Tunable-Mag-order-2022,ACS-Nano-Storchak-EuSi2-2021,Nanoscale-Horiz-Storchak-EuX-2023,Materials-Spectroscopy-TbSi2-2021,Nanoscale-GdGe2-2023} demonstrating the strong 2D ferromagnetic (FM) ordering in the monolayer (ML) limit \cite{Materials-Horizons-Gd-Janus-Wang2020}, high carrier mobility \cite{JMST-high-mobility-Storchak-2023}, non-trivial topology \cite{Adv_Fun_Mat_Topology-Storchak-2020} and remarkable transport phenomena \cite{PhysRevApplied.11.064047}.

The use of materials in electronic and spintronics applications often requires functionalization \cite{Chemical-Reviews_2D-2022,Adv-Sci-Functionalization-2D-Mat-2019,Adv-Funct-Mat-2D-Xenes-2021}. Controlling the electronic and magnetic states in 2D magnetic materials through coupling to external perturbations such as strain \cite{PRB-GdSi2-Cirachi-2021}, stacking \cite{JMCC_GdSGe2-Tunable-Mag-order-2022,Science-stacking-dependent-mag-2019,Yang2022PCCP,ACS-Nano-NiSi2-2021}, formation of Janus heterostructures \cite{Materials-Horizons-Gd-Janus-Wang2020,PCCP-Janus-GdSSe-2023}, pressure \cite{Nat-Mat-pressure-Mag-019}, gating \cite{Nat-Nanotech-gating-CrI3-2018}, electrostatic doping \cite{Nat-Nanotech-2D-mag-Electric-Control-2018,Nat_Rev_Phys-Elect-Field-Mag-2019} and proximity \cite{Nat-Mat-Proximity-2020} became a successful strategy to create novel phases \cite{Mat-Today-Adv-Si-2022}.
%The appropriate implementation of perturbations in the 2D limit offers new opportunities for the creation of quantum phases.
The embedding of Co in graphene activates room-temperature ferromagnetism \cite{Nat-Com-Co-Graphene-2021}; intercalation of Sn in thallene induces emergent spin-polarized states \cite{Mat-Today-Adv-Thallene-2023}, while decoration of germanene \cite{Advanced-Materials-Germanene-2014} with Pb atoms produces the effect of a quasi-freestanding layer with enhanced electronic properties \cite{PhysRevResearch-Pb-Germanene-2021}.
Several recently achieved interface modifications have demonstrated the ability to enhance the stability and enrich the functionalities of 2D-Xenes, thus promoting the rational design of new low-dimensional functional materials \cite{Nano-Letters-Roadmap-2D-2021}.
Recently, a pure antiferromagnetic (AFM) order was reported for \textit{in-situ} grown 2D magnetic bilayers \cite{Nanoscale-GdGe2-2023}, as compared to FM order fingerprints observed in SiO$_x$ capped films \cite{Nat-Comm-Storchak-2D-mag-2018,tokmachev2021two,averyanov2022exchange}, thus demonstrating the importance of taking into account the effects of capping.

In this work, we carry out a combined experimental and theoretical study to investigate the effect of interface modification on the electronic and magnetic properties of the Ln$X_2$ (Ln = Eu, Gd, Dy; $X$ = Ge, Si) family. We discovered that depending on the lanthanide and host elements, a capping layer (used for the prevention of surface degradation in air in earlier reports \cite{Tokmachev2019MH,Parfenov2019} is capable of triggering significant changes in electronic structure and magnetic anisotropy. The detailed descriptions of samples synthesis, experimental measurements, and simulations are given in Appendix, Section 1.

\section{Experimental and calculation details}
\subsection*{Samples growth}
The Ge(111) and Si(111) substrates were used for the growth of the metalloxene films. Ge(111) substrates were sputtered with {Ar+} ion bombardment and then annealed at 650$^{\circ}$C; this procedure was repeated several times until the appearance of sharp c(2$\times$8) low-energy electron diffraction (LEED) pattern. To prepare the Si(111)-7$\times$7 surface reconstruction, Si(111) sample was flash annealed to a temperature of $\sim$1200~$^{\circ}$C. Rare-earth elements were deposited using electron bombardment sources with rates of  $\sim$0.25 ML/min. %[1 monolayer (ML) = 6.2$\times$10$^{14}$ cm$^{-2}$ in terms of the Ge(111) surface atomic density].
The evaporation rate was calibrated by observation of LEED patterns that correspond to known surface reconstructions: 5$\times$2 at coverage less than 1 ML, 1$\times$1 that correspond to the completion of the first ML at coverage $\sim$1 ML and  $\sqrt{3}\times\sqrt{3}$ at coverage above 1 ML \cite{ENGELHARDT2006755}. During deposition, the substrates were held at $\sim$400~$^{\circ}$C. In the case of Si-based metalloxenes, samples were also annealed at $\sim$550-650~$^{\circ}$C to improve the crystalline order of the films. It should be noted that this procedure produces films with multiple film thicknesses after 1 ML completion \cite{ENGELHARDT2006755, Wanke2009SS}.

\subsection*{ARPES and LEED measurements}
The experiments were carried out at the VUV-Photoemission beamline at Elettra synchrotron (Trieste, Italy), using angle-resolved photoelectron spectroscopy (ARPES) and low-energy electron diffraction (LEED) methods. The base pressure of the analytic and preparation chambers was $\leq$ 1.0$\times$10$^{-10}$ Torr and $\leq$ and 3$\times$10$^{-10}$ Torr, respectively. The majority of ARPES measurements were carried out at a temperature of 14 K sing Scienta R4000 electron analyzer and 35 eV photon energy, which allowed for the best contrast of the features of interest. The electron spectrometer was placed at 45$^{\circ}$ with reference to the direction of the incoming \textit{p}-polarized photon beam. The labels of the high-symmetry points in the ARPES spectra refer to the 1$\times$1 surface Brillouin zone (SBZ).

\subsection*{DFT Calculations}
The calculations were based on density-functional theory (DFT) as implemented in the Vienna \textit{ab initio} simulation package VASP \cite{VASP1}. The projector-augmented wave approach \cite{PAW} was used to describe the electron-ion interaction. The generalized gradient approximation (GGA) of Perdew, Burke, and Ernzerhof (PBE) was used as the exchange-correlation functional for structural optimization.  The scalar relativistic effect and the spin-orbit coupling (SOC) were taken into account.

To simulate the LnX$_{2}$ structures, we used a germanium/silicon slab with PBE-optimized bulk lattice constants. Hydrogen atoms were used to passivate the dangling bonds at the bottom of the slab. The kinetic cutoff energy was 250 eV and 12$\times$12$\times$1 and 7$\times$7$\times$1 $k$-point meshes were used to sample the 1$\times$1 and $\sqrt{3}\times\sqrt{3}$ 2D Brillouin zones, respectively. Geometry optimization was performed until the residual force on the atoms was less than 10 meV /\AA. The Heyd-Scuseria-Ernzerhof (HSE06) screened hybrid functional \cite{HSE} was used to calculate the band structure. Two types of Ln pseudopotentials were used \cite{VASP2}: in non-magnetic calculations the potentials where valence 4$f$ electrons are treated as core states were used; while to describe the magnetic properties, standard Ln potentials were used, in which the 4$f$ electrons are treated as valence states.  Since the Gd-based metalloxenes received the main focus of the study, we performed a low-temperature ARPES measurement (14 K) of a thick (20 ML) GdGe$_2$ film to provide experimental evidence for Gd-$f$ band position. As follows from Fig.~S1(a), a strongly localized $f$-band is located at $\sim$-8.5 eV. This observation was used to verify the accuracy of the HSE06 hybrid in the description of the electronic spectra.

\section{Results and discussion}
\subsection{Atomic structure of Ln$X_2$ metalloxenes: the geometry, stoichiometry and stacking type}

\begin{figure*}[t!]
\includegraphics[width=1.0\textwidth]{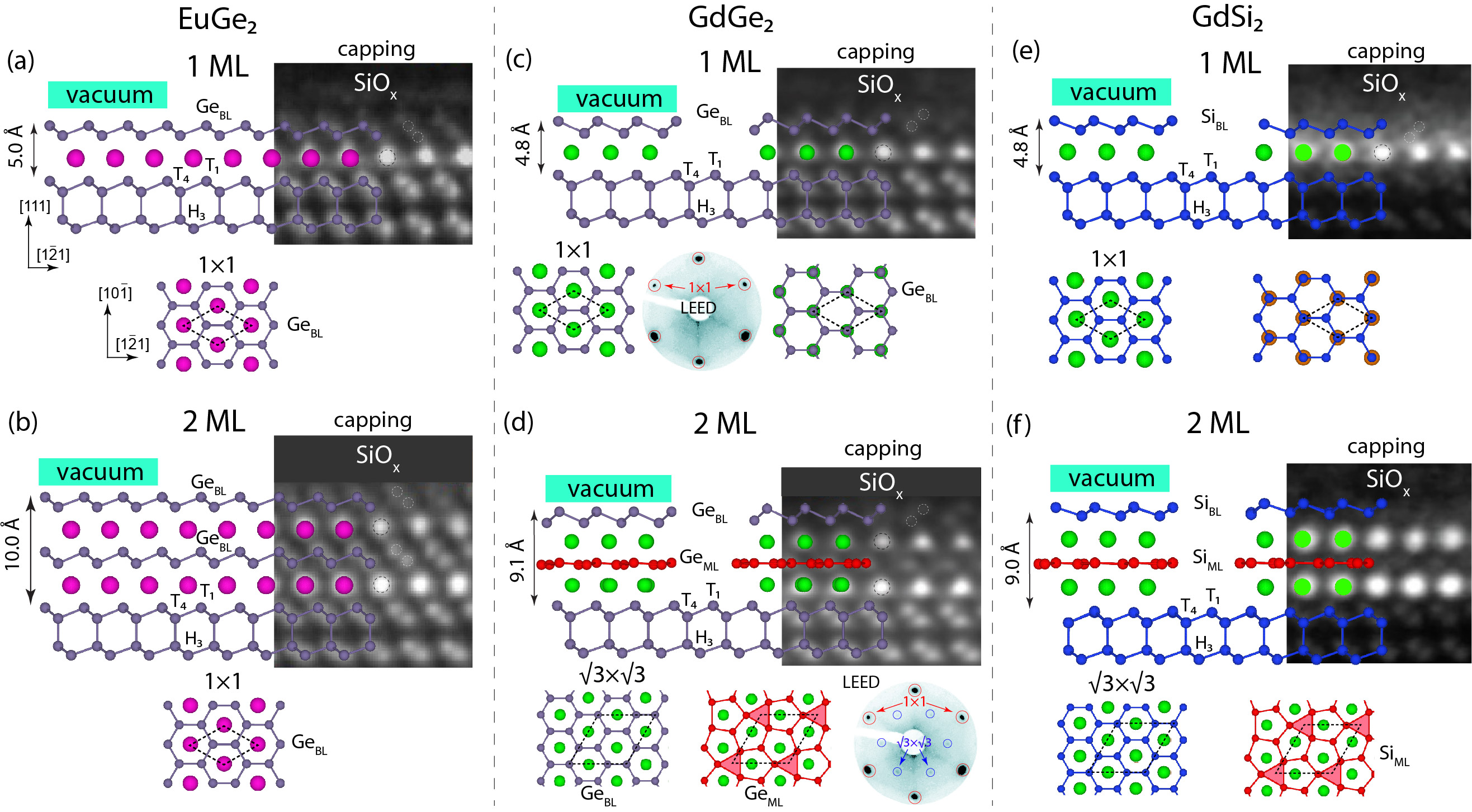}
    \caption{Stacking-dependent ground-state structural models of the EuGe$_2$ (a,b), GdGe$_2$ (c,d) and GdSi$_2$ (e,f) films of various thicknesses. Black dashed rhombi outline the 1$\times$1 (a,b,c,e) and $\sqrt{3}\times\sqrt{3}$ (d,f) unit cells. LEED patterns measured for the 1 ML and 2 ML GdGe$_2$ films are shown in panels (c) and (d), respectively, and demonstrate the reflexes of 1$\times$1 and $\sqrt{3}\times\sqrt{3}$  outlined by red and blue circles, respectively. The observation of the stacking order under the capping layer (a-d) is provided by TEM images reproduced from Ref.~\onlinecite{Tokmachev2019MH} with permission from the Royal Society of Chemistry, and TEM images in (e,f) are reproduced from Ref.~\onlinecite{Parfenov2019} with permission from Elsevier Ltd.}
 \label{Fig-Model}
\end{figure*}

The basic building block of the Ln$X_2$ metalloxene film as follows from high-angle annular dark-field transmission electron microscopy (HAADF-TEM) observations \cite{Tokmachev2019MH} is a triangular lattice of lanthanide ions coupled with honeycomb networks of host Si(Ge) atoms. To theoretically interpret the experimentally observed atomic ordering of the Ln$X_2$ films, we employed the \textit{ab initio} random structure searching (AIRSS) method \cite{Pickard2011} and independently found ground-state models for films of 1 ML and 2 ML thickness. Calculations for EuGe$_2$ show that Eu atoms of the first and second layers prefer to stay at the H$_{3}$ site in reference to the underlying Ge(111) substrate, while the geometry of the segregated buckled honeycomb Ge atomic sheet is the following: the upper Ge atom stays at the T$_{1}$ site, while the lower Ge atom is located at the T$_{4}$ site (Fig.~\ref{Fig-Model}(a)). As regards the 2 ML EuGe$_2$ system, the most energetically stable model has the same arrangement as the 1 ML system: the subsequent Eu and Ge atomic layers follow the same geometry and position with respect to the former ones (Fig.~\ref{Fig-Model}(b)). As a result, the Eu atoms are always centered in the hole of the honeycomb lattice of the lying above Ge bilayer, as shown in the top and side views in Figs.~\ref{Fig-Model}(a,b). %The total thickness of the 1 ML and 2 ML EuGe$_2$ films corresponds to 5.0 \AA\ and 10.0 \AA, respectively.
As found, all theoretical findings appear to perfectly follow the HAADF-TEM observations for EuGe$_2$ \cite{Tokmachev2019MH} shown as the underlying images in Figs.~\ref{Fig-Model}(a,b). This agreement confirms the validity of the AIRSS method and the accuracy of the implemented density-functional theory (DFT) calculation scheme. It should be noted that for the HAADF-TEM measurements the samples were preliminary covered with a 20 nm SiO$_x$ capping layer \cite{Tokmachev2019MH}.
However, excellent agreement in the structure between calculation and experiment allows us to conclude that the capping does not affect the EuGe$_2$ film atomic stacking.

%=======================================================
%=======================================================

Let us now examine the case of GdGe$_2$ films. Figure ~\ref{Fig-Model}(c) shows the ground-state atomic model of the 1 ML film with stacking order as obtained from the AIRSS method; here, the Gd atom, in contrast to Eu, has a different adsorption position -- T$_{4}$ one instead of the H$_{3}$ site for Eu. Next, the geometry of the topmost buckled honeycomb Ge atomic sheet is also different: the upper Ge atom is located at the T$_{1}$ site (which is the same as in the EuGe$_2$ case), but the lower Ge atom is now located in the H$_{3}$ site (instead of the T$_{4}$ site in the EuGe$_2$ case) \cite{Nanoscale-GdGe2-2023}. However, the Gd atoms are positioned at the centers of the honeycomb buckled germanene lattice, like in the case of Eu, which is seen in the top-view atomic model (Fig.~\ref{Fig-Model}(c)). The structure has 1$\times$1 periodicity, which follows from the low energy electron diffraction (LEED) pattern measured for the pristine film under vacuum conditions. Now, if we turn to the HAADF-TEM observations \cite{Tokmachev2019MH} of the 1 ML GdGe$_2$ sample covered by the capping layer, we may see perfect agreement for the position of the Gd atom; however, the stacking of the upper Ge bilayer is different. As follows from the TEM image, the upper Ge atom is located just above the Gd atom at the T$_{4}$ site, while the lower Ge atom stays at the T$_{1}$ site, both different from those in the AIRSS-derived ground state model. So one may conclude on the presence of a lateral shift or slippage of the Ge bilayer relative to the underlying Gd layer. The total energy of the model constructed from the TEM data is 1 eV per unit cell higher than that of the ground-state model. The top-view ball-and-stick atomic model placed just below the HAADF-TEM image in Fig.~\ref{Fig-Model}(c) additionally illustrates capping-induced stacking, where one may see that Gd atoms are located just below one of the Ge sites, but not in the honeycomb centers.

It is worth noting that the SiO$_x$ capping layer was omitted from considerations in our DFT band-structure calculations by the following reasons: SiO$_x$ is a a wide-gap insulator, whose states do not overlap with the LnX$_2$ spectral features of interest; SiO$_x$ is an amorphous phase \cite{Nat-Comm-Storchak-2D-mag-2018} and can not be directly simulated in DFT; and finally, according to the report \cite{Materials-Today-GdSi2-2019} the chemical reaction between SiO$_x$ and LnX$_2$ is unlikely. Therefore, the study of the mechanisms underlying behind the capping-induced changes of the stacking is out of the focus of the present paper.

From a thermodynamic perspective, the GdGe$_2$ film with pristine stacking order is self-sustainable (see phonon calculations in the Appendix, Section 2), but the model derived from the TEM observation shows the instability. Nevertheless, in the presence of a capping layer, the structure becomes stable, as demonstrated in the HAADF-TEM experiments \cite{Tokmachev2019MH,Parfenov2019}.

It is known that the formation of GdGe$_2$ films thicker than 1 ML reduces the symmetry of the surface from 1$\times$1 to $\sqrt{3}\times\sqrt{3}$ \cite{Wanke2009SS}, as demonstrated by the emergence of weak $\sqrt{3}\times\sqrt{3}$ reflexes in the LEED pattern (Fig.~\ref{Fig-Model}(d)). The distinctive feature of multilayer GdGe$_2$ films is the flat intermediate Ge monolayer where the buckling is removed due to the formation of a vacancy \cite{Wetzel1996}. Figure~\ref{Fig-Model}(d) shows the 2 ML ground-state model obtained from our AIRSS calculations, where the intermediate flat Ge ML is marked by red balls, while the vacancies are highlighted by red triangles. The stacking order at the interface between the upper Ge-BL and Gd layer here is the same as in the case of the 1 ML GdGe$_2$ film. As follows from the TEM image, capping the 2 ML GdGe$_2$ sample induces the formation of a different stacking order for the upper Ge-BL (Fig.~\ref{Fig-Model}(d)), alike in the 1 ML GdGe$_2$ film. However, the positions of Gd atoms and atoms of the intermediate flat Ge ML are identical in both films: capped by a protective layer, and the pristine one, determined by the AIRSS calculations.

The structure of the GdSi$_2$ system is almost identical to GdGe$_2$ (Fig.~\ref{Fig-Model}(e)) and the capping-induced stacking order is also less favorable than the pristine ground state (by 0.66 eV). Regarding the DyGe$_2$ system, there is no available HAADF-TEM observation here; nevertheless, DFT calculations (Appendix, Section S5) and angle-resolved photoemission spectroscopy (ARPES) measurements (Appendix, Section S4) suggest that the arrangement of Dy layers and Ge-BL stacking order are identical to the GdGe$_2$ case. Comparison of divalent Eu-based and trivalent Gd(Dy)-based metalloxenes generally shows three distinctive features of the atomic structure: (1) different positions of the Ln atom with respect to the substrate, H$_{3}$ vs. T$_{4}$; (2) buckled honeycomb vs. flat intermediate Ge(Si) layers with vacancy; (3) absence of any changes in the $X_2$/Ln interface vs. modification of the interface stacking under the influence of the capping layer.

\subsection{Stacking- and thickness-dependent band structure of Ln$X_2$ metalloxenes family}

%=======================================================
\begin{figure*}[t!]
\includegraphics[width=1.0\textwidth]{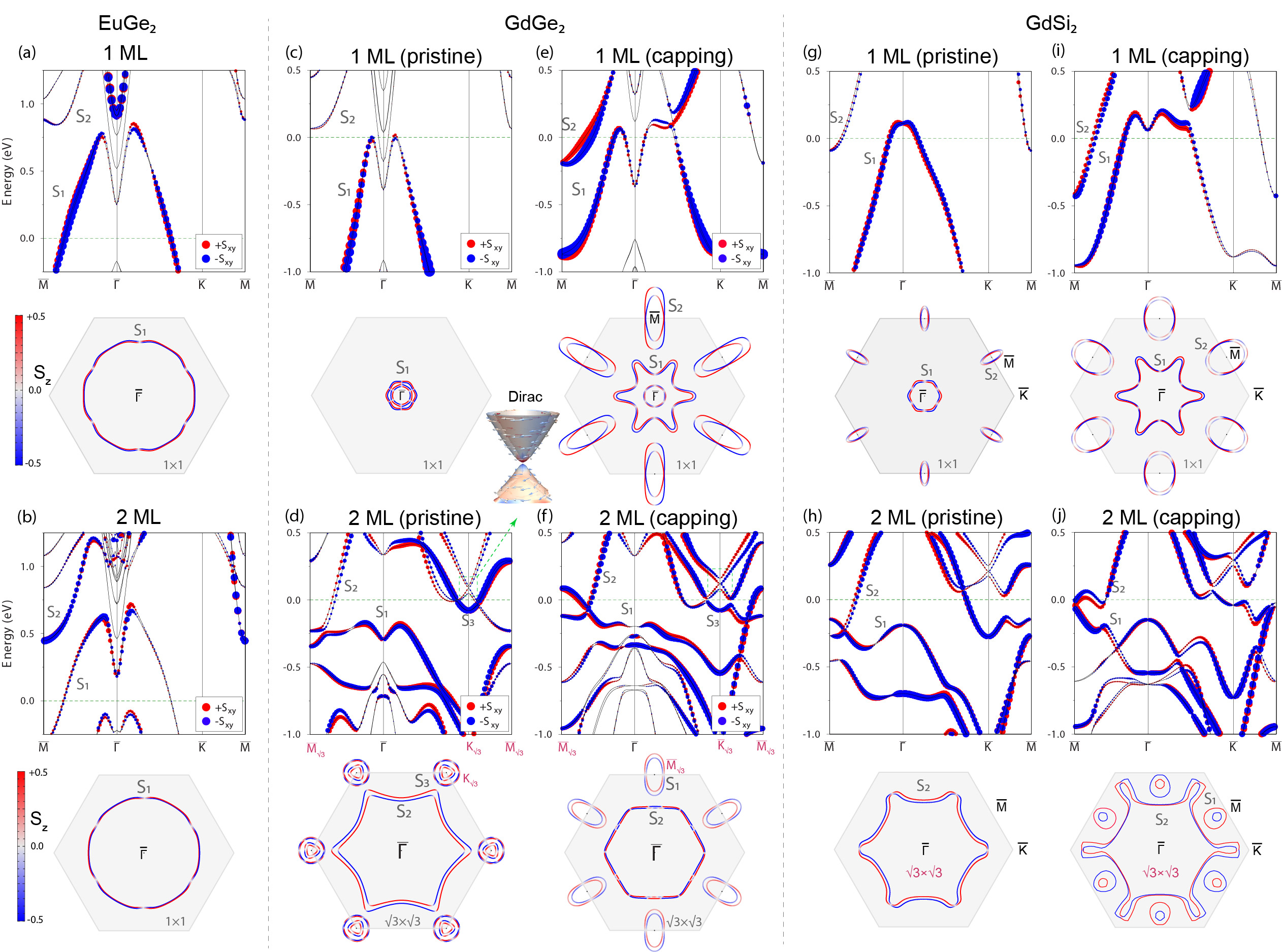}
    \caption{Spin-resolved electronic spectra and Fermi-contour maps calculated for (a,b) EuGe$_2$, (c-f) GdGe$_2$ and (g-j) GdSi$_2$ films of various thickness and stacking type. Red and blue in the electronic spectra and Fermi maps represent opposite in-plane and out-of-plane components of the spin expectation values, respectively, while the balls' size reflects the magnitude of the spins. The highlighted hexagons denote the first Brillouin zone. The image above (d) shows an enlarged view of the Dirac-like cones in the unoccupied spectrum.}
 \label{Fig-BND+Fermi-mapping}
\end{figure*}
%=======================================================

Let us now consider the electronic properties of EuGe$_2$ films, whose structure does not experience any changes induced by the capping layer. Figure~\ref{Fig-BND+Fermi-mapping}(a) shows the spin-resolved relativistic electronic spectrum of the 1 ML EuGe$_2$ film in the non-magnetic case. The distinctive feature of the spectrum is the highly dispersed in-gap S1 hole-like metallic band that possesses (small) Rashba-type spin splitting and demonstrates hybridization with the Ge(111) conduction band at the $\bar\Gamma$ point. Another distinctive feature is an almost degenerate unoccupied in-gap S2 band with electron-like dispersion at the $\bar{\textrm{M}}$ point. The Fermi-surface map shows $\bar{\Gamma}$-centered almost isotropic hole pocket that demonstrates insignificant spin splitting. The spectrum of the 2 ML EuGe$_2$ film has very similar features -- the S1 band (which intersection with the Fermi level ($E_{\mathrm F}$) produces a pair of circular contours) and the parabolic S2 in-gap band dispersing by 0.5 eV above the $E_{\mathrm F}$. In general, the 1 ML and 2 ML EuGe$_2$ films belong to the class of $p$-type metals.

Figure~\ref{Fig-BND+Fermi-mapping}(c) shows the DFT calculated band structure of 1 ML  GdGe$_2$ film with the pristine stacking, and at first glance it has a very similar character to its EuGe$_2$ counterpart, namely two in-gap bands, the S1 highly-dispersing metallic state and fully unoccupied S2 band. However, the position of the Fermi level here is 0.75 eV higher than in the EuGe$_2$ case and is pinned just at the minimum of the Ge-bulk conduction band. As follows from Fig.~\ref{Fig-BND+Fermi-mapping}(c) the Fermi surface map has a tiny hole-like pocket produced by the S1 band and even a much smaller electron-like pocket produced by the bulk states, which all characterize the system as a compensated weak metal. Figure~\ref{Fig-BND+Fermi-mapping}(e) shows the spectrum of the GdGe$_2$ film with staking corresponding to the structure with capping layer. As one can see, the change in stacking order produces drastic changes in the band structure. The most appealing effect is the significant enhancement of the electron density around the Fermi level, provided by the S2 electron band that forms elongated $\bar{\textrm{M}}$-centered electron pockets with significant Rashba spin splitting in the $\bar{\Gamma}$-$\bar{\textrm{M}}$ direction. The Fermi level position was also changed, and now it is placed within the Ge bulk gap. One may see that capping-induced changes in the atomic structure significantly enrich the electronic properties of the 1 ML GdGe$_2$ that becomes a compensated metal with dominant $n$-type charge carriers.

Let us now consider the electronic spectra of the 2 ML GdGe$_2$ films with pristine and capping-induced geometry shown in Fig.~\ref{Fig-BND+Fermi-mapping}(d) and Fig.~\ref{Fig-BND+Fermi-mapping}(f), respectively. In contrast to EuGe$_2$ case the band structure of the 2 ML GdGe$_2$ film substantially differs from that of the single layer. In the case of a pristine film, the Fermi level is placed exactly in the middle of the Ge bulk gap, whereas S2 and S3 metallic bands form $\bar{\Gamma}$-centered and $\bar{\textrm{K}}$-centered pockets, respectively, with more noticeable spin splitting compared to the 1 ML film.
One may see that the capping-modified interface experiences substantial changes in the electronic spectrum: the S1 band becomes metallic and forms a pair of $\bar{\textrm{M}}$-centered hole-like pockets, while the S3 band does not produce any pockets at $E_{\mathrm F}$ due to the hybridization gap. It should be noted that application of a positive perpendicular electric field to the GdGe$_2$ system can easily move the S1 spin-split band at the Fermi level and increase the density of states at $E_{\mathrm F}$ (see Appendix, Section S6).

Now, we trace the effect of lanthanide and host element variation on the electronic properties of metalloxene. Section~S3 in Appendix summarizes the results for DyGe$_2$ films, whose atomic and electronic band structure are almost identical to those of GdGe$_2$. Figure~\ref{Fig-BND+Fermi-mapping}(g-j) shows the results for 1 ML and 2 ML GdSi$_2$ films that show a stronger energy overlap between the states at $\bar{\Gamma}$ and at $\bar{\mathrm M}$ as compared to GdGe$_2$. In the Fermi surface, this results in an appearance of additional pockets at $\bar{\mathrm M}$ for 1 ML and the absence of circular features at $\bar{\mathrm K}$ for 2 ML, characterizing 1 ML and 2 ML films in the pristine case as compensated and $p$-type metals, respectively. Band structure for the capped 1ML  GdSi$_2$ show similar to the 1 ML GdGe$_2$ case appearance with the Fermi surface that consists of star-like hole pocket at $\bar{\Gamma}$ point and elongated pockets in $\bar{\mathrm M}$ points. In case of capped 2 ML GdSi$_2$ film the hybridization effects distort the S2 band and lead to a doubling of the $\bar{\mathrm M}$-centered pocket.
Interestingly, the pristine and capped 2 ML GdGe(Si)$_2$ films demonstrate a Dirac-like cone at the $\bar{\mathrm K}$ point just above the Fermi level (shown as the outset above panel (d)), which is a typical feature of a 2D-Xene honeycomb material. ARPES observations for the \textit{in-situ} grown GdGe$_2$, GdSi$_2$ and DyGe$_2$ films are in very good agreement with the DFT predictions for pristine films (see Ref.~\onlinecite{Nanoscale-GdGe2-2023} and Appendix, Section~S4, respectively), confirming that the \textit{in-situ} grown films have a different structure as compared to the SiO$_x$-capped.

%=======================================================
\begin{figure*} [t!] % [htbp]
    \includegraphics[width=1.0\textwidth]{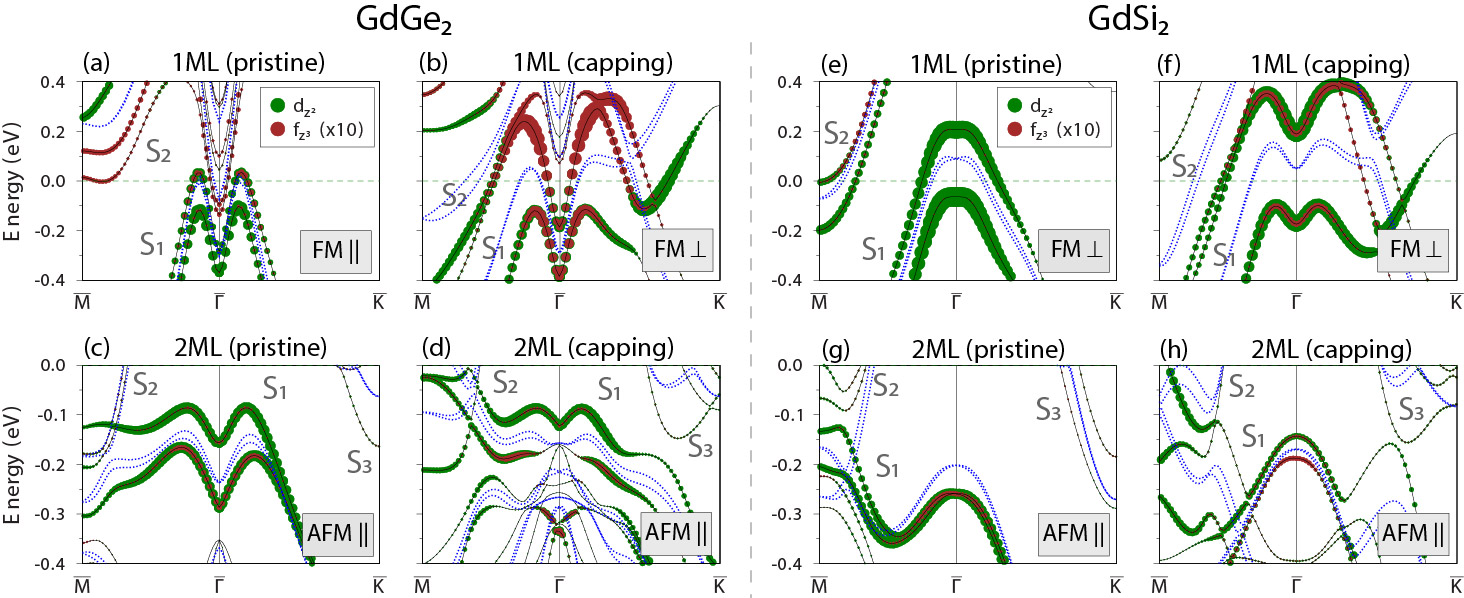}
    \caption{The effect of magnetism on the electronic band structure of GdGe$_2$ (a-d) and GdSi$_2$ (e-h) films with pristine and capping-induced stacking orders. The spectra demonstrate the contribution of the Gd-$d$ and Gd-$f$ orbitals to the electronic structure. The blue dashed and black solid curves demonstrate the non-magnetic and magnetic spectra, respectively. The preferred magnetic configuration for each system is shown on the panels.}
 \label{Fig-BND-magnetic}
\end{figure*}
%=======================================================

\subsection{Stacking-dependent magnetic order of LnX$_2$ metalloxenes}

The profound effect of the stacking order on the band structure dispersion (Fig.~\ref{Fig-BND+Fermi-mapping}) and orbital character (Appendix, Section S6) of the GdGe(Si)$_2$ films must leave an imprint on the magnetic ground state and consequently on their electronic band structure in the magnetic state. To this aim, first of all, we define the preferred magnetization direction by calculating the total energies for various magnetic configurations (Appendix, Section S7); we found that for 2 ML films the AFM$_\|$ is preferred magnetic configuration for both systems. However, for 1 ML films, the magnetic ground state depends on the Ln element and the stacking type: for pristine stacking it is FM$_\|$ for GdGe$_2$, whereas it is FM$_\bot$ for the GdSi$_2$.
Next on the basis of preferred magnetic configuration we calculated the band structure. Figure~\ref{Fig-BND-magnetic} shows the orbital-resolved and stacking-dependent band structures of the magnetic GdGe$_2$ and GdSi$_2$ films of 1 ML and 2 ML thickness. To trace the dispersion changes, we show the non-magnetic spectra as blue dashed curves. The band structure of the pristine 1 ML GdGe$_2$ film (Fig.~\ref{Fig-BND-magnetic}(a)) demonstrates a noticeable exchange splitting in both the S1 and S2 surface bands, with lifting the degeneracy at the $\bar\Gamma$ and $\bar{\rm M}$ points. Remarkably, in the case of the capped 1 ML film (Fig.~\ref{Fig-BND-magnetic}(b)) the amplitude of exchange splitting is much larger than in the pristine case. Orbital-symmetry analysis shows that the exchange splitting of S1 and S2 bands comes from Gd $f-d$ hybridization, mediated through the out-of-plane $f_{z^3}$ and $d_{z^2}$ components (shown in Fig.~\ref{Fig-BND-magnetic} as brown and green circles), which acquire larger weights in the case of a capped film. The exchange splitting in the spectrum of the 2 ML GdGe$_2$ film is rather large in both the pristine and capped cases. As for the case of pristine GdSi$_2$ system (Fig.~\ref{Fig-BND-magnetic}(e)) one may see a significant exchange splitting already at the monolayer limit in contrast to the Ge-based counterpart; however, the capping-induced modification of the film stacking hugely enhances the band splitting, which is again associated with the increased contribution of $f_{z^3}$ orbitals to the S1 and S2 bands (see Fig.~\ref{Fig-BND-magnetic}(e)).
Therefore, the study of magnetism-induced effects on the electronic band structure of Ln$X_2$ metalloxenes shows the pivotal role of Ln $f-d$ hybridization \cite{AdvElMat-Ce-f-d-Vyalikh-2022,Nanoscale-CePb3-2022}, which can be effectively tuned by modification of the interface. The increase in film thickness up to 2 ML leads to a substantial decrease in the exchange splitting in the electronic spectra of the GdSi$_2$ films regardless of the type of stacking order due to anti-parallel alignment of the magnetic moments in the neighboring layers.
The DFT predictions for the AFM order and the corresponding band splitting in 2 ML GdGe$_2$ films are in line with ARPES results (Ref.~\onlinecite{Nanoscale-GdGe2-2023}, Fig.~\ref{Fig-ARPES-magnetism} in Appendix). Due to a lower temperature of the transition, we were not able to confirm such a splitting in 1 ML GdGe$_2$. Similarly, we were not able to observe any temperature-dependent splitting in other Ln$X_2$ systems. Since the EuGe$_2$ films do not experience any stacking modification induced by the capping layer, we give only a short discussion for them (see Section S8, Appendix). As regards the case of the EuSi$_2$ and DyGe$_2$ systems, we do not have any (HAADF-TEM) evidence on the effect of the capping layer and therefore leave them out of consideration.

\section{Conclusions}
In summary, we reveal that external perturbations, such as film capping, trigger profound changes in the electronic structure and magnetic ground state of the ultra-thin Ln$X_2$ metalloxenes, which is unveiled by DFT calculations combined with ARPES observations. The comprehensive characterization of the atomic structure provided by the earlier HAADF-TEM observations and our \textit{ab initio} random structure searching calculations allowed us to identify the geometry of metalloxenes formed in the presence of a capping layer and without it. It is found that capping of EuGe$_2$ films does not produce any structural changes in the geometry of the germanene stacking, while the surfaces of the GdGe$_2$ and GdSi$_2$ films are substantially modified, which inadvertently affects the electronic states and magnetic order. As such, the pristine 1 ML GdGe$_2$ film is characterized as a weak $p$-type metal, while the electronic spectrum of the capped 1 ML GdGe$_2$ film has significant electron density at the Fermi level and around, demonstrating characteristics of a compensated metal. Variation of the host element also produces substantial changes in the electronic structure, namely, in the GdSi$_2$ films the increased concentration of charge carriers is predicted. Modification of metalloxene stacking was found to trigger a change in magnetic order: in the 1 ML GdGe$_2$ film it was changed from FM$_\|$ to FM$_\bot$. Additionally, we find that an external electric field may be used as an effective tool for tuning the position of the Fermi level. Thus, our study offers insights into the electronic properties of the 2D ultrathin rare-earth metalloxenes and their effective functionalization.

\appendix
\section*{Appendix}

\subsection{ARPES observation and DFT prediction of 4$f$ states position in GdGe$_2$ film.}

\begin{figure}[h!]
\centering
\includegraphics[width=\columnwidth]{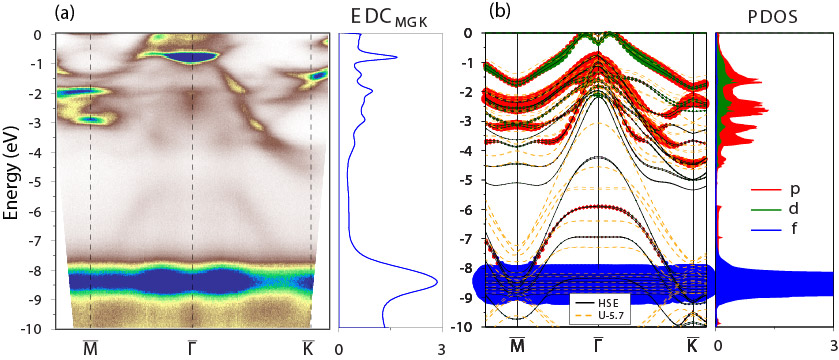} 
 \caption{(a) Large-energy scale ARPES spectrum of the thick GdGe$_2$ film (20 ML) demonstrating the position of Gd-$f$ band (spectrum is taken at 14 K, with 35 eV photons and recorded within 1$\times$1 SBZ). The angle-integrated energy distribution curve (EDC) defines the position of the Gd-f states at -8.5 eV, while multiple Gd-d and Ge-p states are between 0 and -4 eV. Orbitals-projected spectrum and partial density of states (PDOS) calculated by HSE06 for the 1 ML GdGe$_2$ film with FM$\bot$ magnetic ordering. The black solid and orange dashed curves demonstrate spectra obtained within HSE06 and DFT+$U$ ($U$=5.7 eV) approaches.}
 \label{Fig-ARPES-f}
\end{figure}

ARPES spectrum of the thick GdGe$_2$ film (20 ML) is shown in Fig.~\ref{Fig-ARPES-f}.

 %=======================================================
%=======================================================

%\clearpage
\subsection{The thermodynamic stability of pristine and capped GdGe$_2$ films.}

\begin{figure}[htbp]
\centering
\includegraphics[width=\columnwidth]{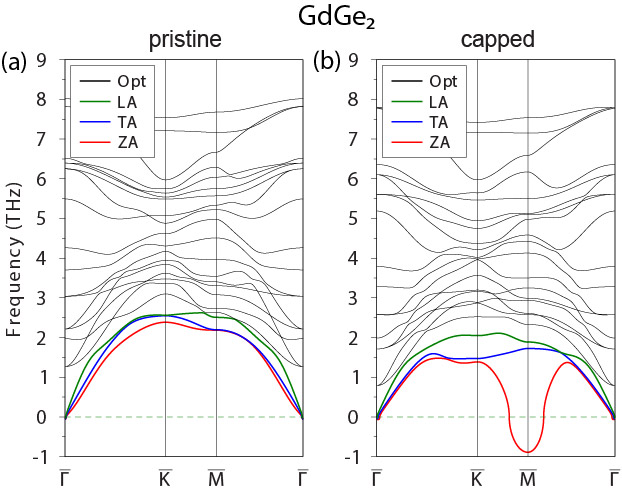}
\caption{Phonon spectra of 1 ML GdGe$_2$/Ge(111) system with (a) pristine and (b) capping-induced stacking order of Ge-BL. The out-of-plane acoustic mode (ZA) is colored red, the transverse acoustic mode (TA) blue, the longitudinal acoustic mode (LA) green, and the optical modes black. }
 \label{fig-Phonons}
\end{figure}

To study the stability of the pristine and capped 1 ML GdGe$_2$ film, we performed phonon calculations. Figure~\ref{fig-Phonons} demonstrates the corresponding vibrational spectra. From the thermodynamic perspective, the structure with pristine stacking order in the GdGe$_2$ film is found to be self-sustainable as follows from the phonon spectra $\omega$$_i$(q) (Fig.~\ref{fig-Phonons}(a)) obtained in the relativistic and non-magnetic approach with the finite-displacement method implemented in phonopy \cite{Scr-Mater-Phonopy-2015}. The vibrational fingerprint of a system with pristine Ge-BL stacking is characterized by three distinct phonon modes (ZA, TA, LA) in which both the energies and symmetries of these modes demonstrate the evident thermodynamic stability. 
However, the phonon spectrum of the structure derived from the HAADF-TEM image for the capped film (Fig.~\ref{fig-Phonons}(b)) demonstrates the emergence of an imaginary mode around the $\bar{\textrm{M}}$ point reaching the -1~THz value, which indicates strong instability of the system and a displacive phase transition. However, despite its intrinsic instability, in the presence of a capping layer, the structure is stable, which is confirmed by HAADF-TEM.

%+++++++++++++++++++++++++++++++++++++++++++++
%+++++++++++++++++++++++++++++++++++++++++++++

\clearpage
\subsection{DFT predictions of DyGe$_2$ films band structure.}

\begin{figure}[h!]
	\centering
 \includegraphics[width=\columnwidth]{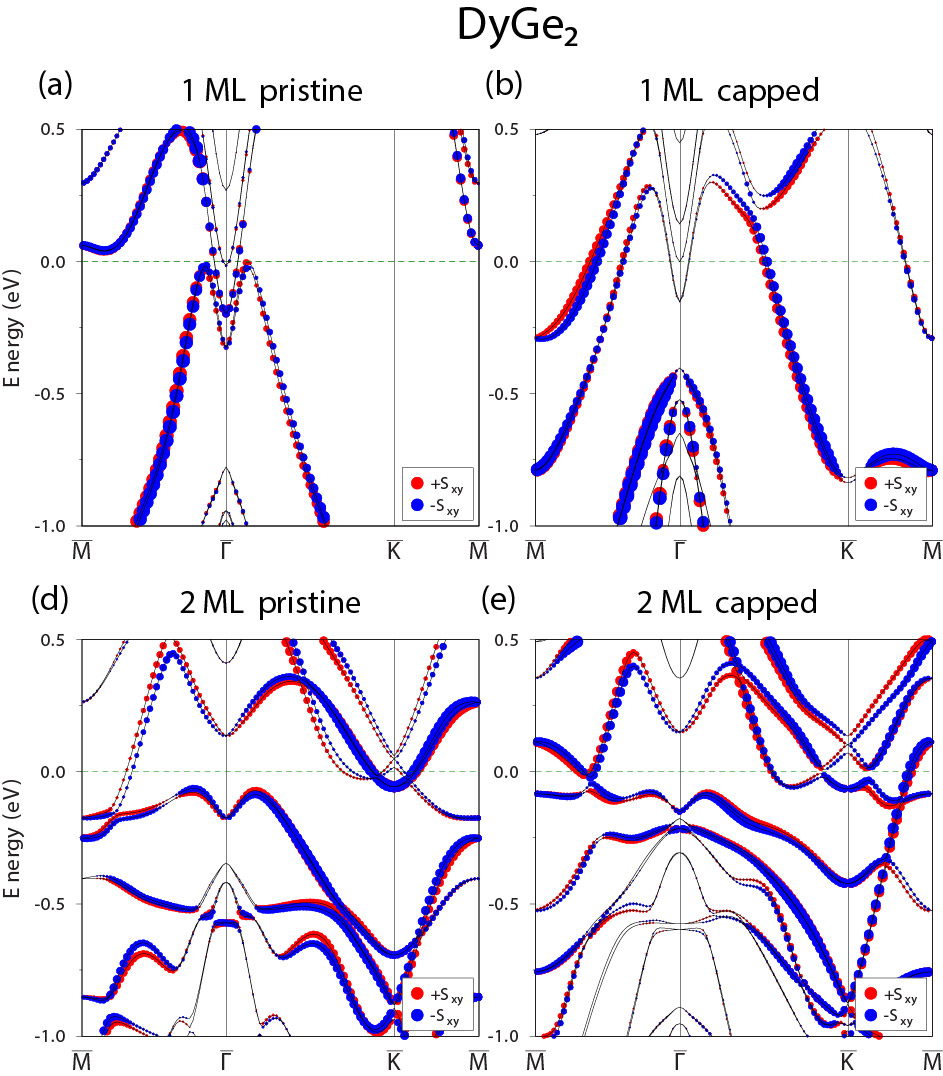}
	\caption{Spin-resolved electronic spectra calculated for DyGe$_2$ films of various thickness and stacking type. Red and blue balls in the electronic spectra represent the direction of the in-plane components of the spin expectation values, while the balls' size reflects the magnitude of the spins.  }
 \label{Fig-ARPES+GdSi-Dy-Ge}
\end{figure}

Spin-resolved electronic spectra calculated for 1 ML and 2 ML DyGe$_2$ films with pristine stacking, and stacking formed under the influence of the capping layer, are shown in Fig.~\ref{Fig-ARPES+GdSi-Dy-Ge}.

%=======================================================
 %=======================================================
%+++++++++++++++++++++++++++++++++++++++++++++

%\clearpage
\newpage
\subsection{ARPES observations of the electronic structure of GdGe$_2$, GdSi$_2$ and DyGe$_2$ films}

\begin{figure}[h!]
	\includegraphics[width=\columnwidth]{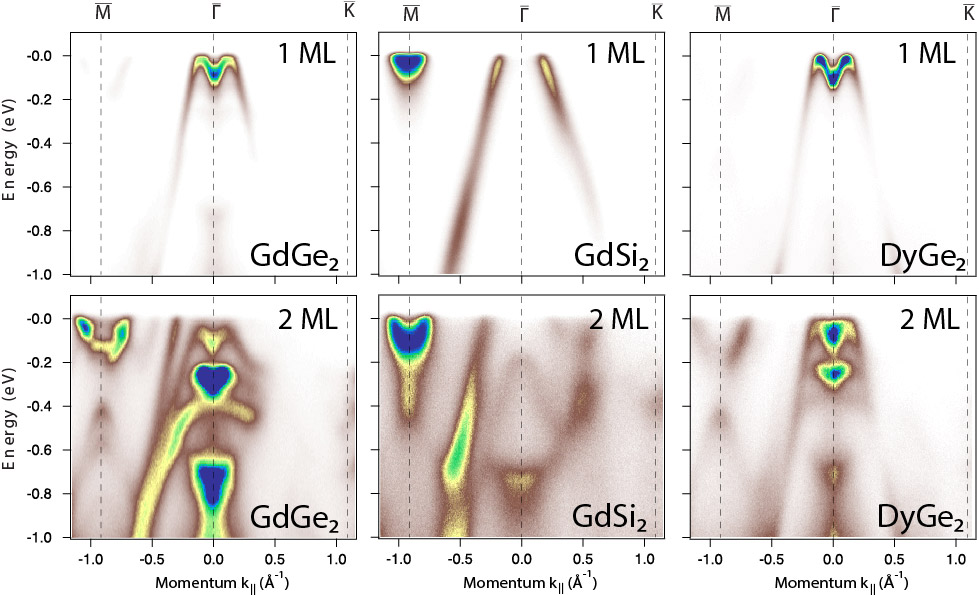}
	\caption{ARPES spectra for 1 ML (upper row) and 2 ML (lower row) GdGe$_2$, GdSi$_2$ and DyGe$_2$ films recorded within 1$\times$1 SBZ. All spectra are measured with 35 eV photon energy, at a temperature of 80 K, which corresponds to a non-magnetic phase. The ARPES data demonstrate the spectral characteristics that nicely correspond to the predicted band structures discussed in the main text, in line with Ref.~\onlinecite{Nanoscale-GdGe2-2023}.}  
 \label{Fig-ARPES-general}
\end{figure}

\begin{figure}[h!]
	\includegraphics[width=\columnwidth]{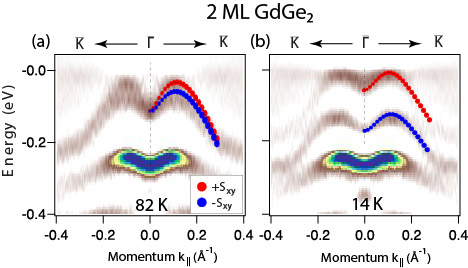}
	\caption{(a,b) Second derivatives of the ARPES electronic spectra in the vicinity of $\bar\Gamma$ point of the 2 TL GdGe$_2$ film taken at temperatures 14 K and 82 K, respectively, with 35 eV photon energy. Spectrum in (a) corresponds to paramagnetic, and spectrum in (b) to magnetic. The schematic spin-polarized bands are overlaid on the experimental data to demonstrate the change of the spectrum with the temperature, in line with Ref.~\onlinecite{Nanoscale-GdGe2-2023}. The red and blue balls represent the opposite in-plane spin components.}  
 \label{Fig-ARPES-magnetism}
\end{figure}

ARPES spectra for 1 ML and 2 ML GdGe$_2$, GdSi$_2$ and DyGe$_2$ films in paramagnetic phase are presented in Fig.~\ref{Fig-ARPES-general}. Second derivatives of the ARPES electronic spectra for the 2 TL GdGe$_2$ film at temperatures 14 K and 82 K are shown in Fig.~\ref{Fig-ARPES-magnetism}.

%=======================================================
%=======================================================

%=======================================================
%=======================================================

%\clearpage
\newpage
\subsection{Influence of external electric field on the band structure of GdGe$_2$ film.}

%===========================================================
\begin{figure}[htbp]
	\includegraphics[width=\columnwidth]{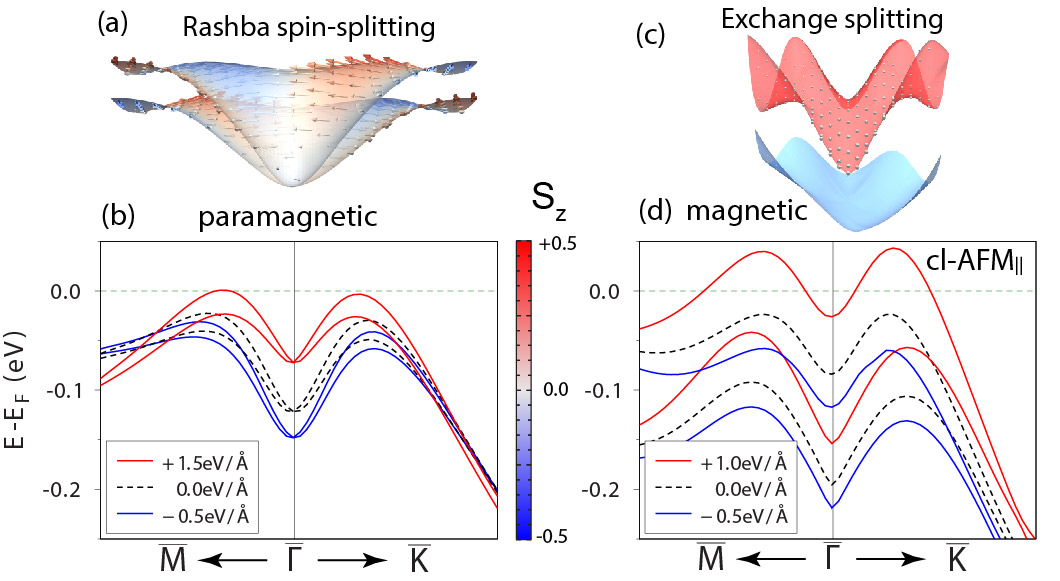}
	\caption{(a) and (c) show the 3D perspective view of the electronic surface bands of 2 ML GdGe$_2$ system and associated spin texture in non-magnetic and magnetic case, respectively. The arrows represent the direction of the in-plane spin components; red(blue) colors mark the positive(negative) sign of the out-of-plane spin component. Panels (b) and (d) show the set of relativistic electronic spectra calculated for non-magnetic and magnetic (collinear-AFM$_{||}$) cases, respectively, under external perpendicular electric field at representative $E$=-0.5, 0.0, +1.0 and +1.5 eV/\AA. }
 \label{Fig-EF}
\end{figure}
%===========================================================

In the current section it is shown that the spin-polarized band dispersing just below the Fermi level in the 2 ML GdGe$_2$ film can be shifted exactly to the Fermi level by the external perpendicular electric field $E$. The 2 ML GdGe$_2$ interface has out-of-plane structural asymmetry, which implies that the application of the $+z$ and $-z$ directional electric field is not equivalent. Our calculations show that the position of the surface band is shifted to the Fermi level by positive $E$, while negative $E$ moves the band down as seen in Fig.~\ref{Fig-EF}. With considered positive $E=+1.5$ eV/\AA\ and $E=+1.0$ eV/\AA\ moves the surface band exactly at the Fermi level for the paramagnetic and magnetic phases, respectively.

%\clearpage
\subsection{Orbital-symmetry analysis of GdGe$_2$ system}
%=======================================================
\begin{figure}[htbp]
\centering
\includegraphics[width=\columnwidth]{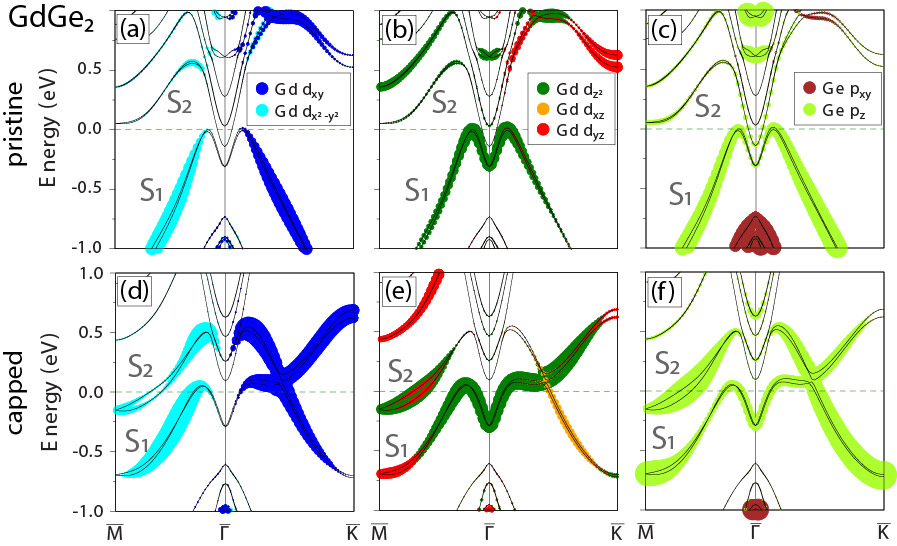}
	\caption{The character of Gd-Ge hybridization within 1 ML GdGe$_2$ film. Orbital-projected relativistic spectra for the pristine (a,b,c) and capped (d,e,f) films demonstrating the distribution of the Gd-d and Ge-p states in the vicinity of E$_F$.}
\label{Fig-BND-orbitals}
\end{figure}
%=======================================================
To provide insights into the nature of changes in the band structure induced by the interface modification, we performed a comprehensive orbital-symmetry analysis and examined the nature of Gd-Ge hybridization.
Figure~\ref{Fig-BND-orbitals} shows the orbital-projected band structures for the pristine (a-c) and capped (d-f) 1 ML GdGe$_2$ films. In the pristine film, the metallic S1 band is formed mainly by the in-plane Gd $d$ orbitals ($d_{xy}$ and $d_{x^2-y^2}$ (Fig.~\ref{Fig-BND-orbitals}(a)), however, in the vicinity of $E_F$ (where the S1 band hybridizes with the bulk states of Ge at the $\bar{\Gamma}$ point) one may see the dominance of the Gd out-of-plane $d$-orbitals, especially $d_{z^2}$. The S2 band demonstrates an orbital composition similar to the S1 band, but with a leading contribution of the $d_{yz}$ states and much smaller orbital weights. As regards the Ge contributions, the S1 and S2 bands are determined solely by the $p_{z}$ orbitals, %(with orbital weights of the S1 band 3 times larger than the S2 weights), 
while the $p_{xy}$ orbitals contribute to the Ge valence band (Fig.~\ref{Fig-BND-orbitals}(c)). 
The main change in the electronic spectrum induced by the capping is the shifting of the S2 band down to the vicinity of the Fermi level. From the perspective of orbital contributions, there are also a number of changes: the $d_{yz}$ and $d_{xz}$ states now are noticeably intermixed with $d_{z^2}$ in both metallic bands at the edges of the Brillouin zone (Fig.~\ref{Fig-BND-orbitals}(e)). The capping also induces an increase in the weights of the out-of-plane Ge $p$ orbitals in the S2 band in the vicinity of E$_F$ (Fig.~\ref{Fig-BND-orbitals}(f)) indicating the increase in Gd-Ge hybridization. 

%\clearpage
\subsection{Magnetic properties of LnX$_2$ metalloxenes}

First we would like to provide a general background on magnetic and transport studies of LnX$_2$ metalloxenes and after to examine the magnetic-induced changes in their electronic band structure.
As reported in Refs.~\cite{Materials-Today-GdSi2-2019,Sokolov2020MH} few monolayer films of GdSi$_2$, EuSi$_2$, and EuGe$_2$ demonstrate a fingerprints of robust 2D FM order with in-plane orientation of the magnetic moments; however, as soon as the films become relatively thick (for example, 17 ML for GdSi$_2$ \cite{Nat-Comm-Storchak-2D-mag-2018} or above), the 2D FM order converts into an A-type 3D AFM structure with in-plane orientation of magnetic moments, formed by FM layers. The Neel temperature for the metalloxene in a form of thick films was found to be 50 K (Gd$_3$Si$_5$)\cite{Tokmachev2018NC}, 38 K (Gd$_3$Ge$_5$) \cite{Tokmachev2019MH}, 48 K (EuGe$_2$)\cite{Tokmachev2019MH}, 41 K (EuSi$_2$)\cite{Averyanov2016}. These values are in line with the older studies on bulk crystals\cite{RE_book}. 
%\hl{I suggest to stop here because it is an SI and not a review on Storchak works (put the rest part under LATEX comment). }
%Regarding transition temperatures, magnetotransport studies have shown that $T_c$ in thin LnX$_2$ films \add{varies}\out{behaves} as a function of the external magnetic field \cite{Nanoscale-Horiz-Storchak-EuX-2023},
 %which is a hallmark of intrinsic 2D FM, when the external field affects the spin-wave excitation (pseudo) gap \cite{Nature-intrinsic-FM-vdW-2017}, enabling long-range FM order at finite temperatures. 
%However since the current study is focused on the intrinsic magnetism of LnX$_2$ metalloxenes, we will list the \hl{define it} $T_c$ values for zero magnetic field. As such, the thick GdSi$_2$ film has $T_c$ = 50 K \cite{Nanoscale-Horiz-Storchak-EuX-2023}, thick GdGe$_2$ film demonstrates $T_c$ = 38 K \cite{Sokolov2020MH}, 9 ML film of EuSi$_2$ has T$_c$ = 80 K \cite{Nanoscale-Horiz-Storchak-EuX-2023} and the thick EuGe$_2$ film has $T_c$ = 38 K \cite{Sokolov2020MH}. Finally, according to the recent report [\cite{Nanoscale-GdGe2-2023}] the $T_c$ for the 2 ML GdGe$_2$ film was estimated to be 31 K, while $T_c$ for the 1 ML film was found to be lower than 14 K. 
As regards the magnetic moments of Ln elements in the ultrathin LnX$_2$ metalloxenes that show fingerprints of the FM behaviour, it is observed that the absolute value of the saturation moment is much lower than the 7 $\mu_{\rm B}$, expected for half-filled 4$f$ shells of Eu$^{+2}$ and Gd$^{+3}$ ions \cite{Nat-Comm-Storchak-2D-mag-2018}. Although the origin of the reduced moment observed experimentally in various magnets is unclear, there are suggestions that it could possibly be related to magnetic fluctuations that arise from magnetic frustration \cite{Nano-Letters-MBS-2021}, structural disorder \cite{Nat-Com-MBT-2020}, non-ionic character of the silicides (germanides) and AFM fluctuations arising from competing magnetic interactions \cite{Nat-Comm-Storchak-2D-mag-2018}.
\begin{table} [htbp]
\begin{center}
\begin{tabular}{|c|c c c|c c|c c c|c c|c|} 
\hline 
\multicolumn{1}{|c|}{Stacking } & \multicolumn{5}{|c|}{pristine} & \multicolumn{5}{|c|}{capping} \\
\hline 
\multicolumn{1}{|c|}{Magnetic order} & \multicolumn{3}{|c|}{AFM} & \multicolumn{2}{|c|}{FM} & \multicolumn{3}{|c|}{AFM} & \multicolumn{2}{|c|}{FM} \\
\hline 
\hline 
 GdGe$_{2}$ & ${||}$   & $\bot$  &  nc & ${||}$   & $\bot$ & ${||}$   & $\bot$  &  nc & ${||}$   & $\bot$ \\
\hline 
1 ML  &    &      &      &  0.0  & 0.2  &    &      &      &  0.4  & 0.0   \\
\hline 
2 ML  &  0.0 &  0.2   & 1.9   & 10.0  &  11.0 &  0.0 &  0.1    & 5.1  & 10.1  &  10.2    \\
\hline
\hline 
 GdSi$_{2}$ & ${||}$   & $\bot$  &  nc  & ${||}$   & $\bot$ & ${||}$   & $\bot$  &  nc  & ${||}$   & $\bot$ \\
\hline 
1 ML  &    &      &      &  1.4  & 0.0  &    &      &      &  1.4  & 0.0   \\
\hline 
2 ML  &  0.0 &  0.1  & 2.1   & 13.1  &  13.2  &  0.0 &  0.1 & 8.0   & 14.0  &  14.5   \\
\hline 
\end{tabular}
\end{center}
\caption{The calculated relative total energy (meV/Gd atom) with respect to the ground state for 1 -- 2 ML GdGe$_{2}$ and GdSi$_{2}$ films with FM and AFM interlayer coupling for pristine and capped films.
[collinear in-plane AFM (AFM$_\|$), collinear out-of-plane AFM (AFM$\bot$); non-collinear 120${}^\circ$ AFM (nc AFM$_\|$), in-plane FM (FM$_\|$), out-of-plane FM (FM$\bot$)].}
\label{tab1}
\end{table}
%+++++++++++++++++++++++++++++++++++++++++++++

As shown in Tab.~S1, the in-plane FM (FM$_\|$) and AFM (AFM$_\|$) are found to be the most energetically favorable magnetic configurations for the pristine 1 ML and 2 ML GdGe$_2$ films, respectively, which is in agreement with experiment \cite{Materials-Today-GdSi2-2019,Sokolov2020MH}. The capped 1 ML GdGe$_2$ film, in contrast to the pristine one, holds the FM$_\bot$ configuration as energetically more favorable, showing that the modification of the stacking leads to a change in magnetic anisotropy. Regarding the case of the 2 ML GdGe$_2$ system, the AFM$_\|$ configuration is estimated here as a ground state; AFM$_\bot$ is only by 0.2 meV (per Gd atom) less favorable, while the non-collinear in-plane AFM (ncl AFM$_\|$), as well as the FM$_\|$ and FM$_\bot$ configurations are much less favorable. 
The change in the host element in GdX$_2$ from Ge to Si induces a change in the magnetic ground state of the 1 ML film and is predicted as FM$_\bot$ for the pristine case. As regards the film with stacking formed under the presence of the capping layer, one may see the keeping of the same AFM$_\|$ magnetic ground state. Finally, the calculations for 2ML GdSi$_2$ films of both stacking orders also reveal AFM$_\|$ as a more preferable configuration.

\subsection{Thickness-dependent electronic band structure of EuGe$_2$ films in magnetic phase}

%=======================================================
\begin{figure}[htbp]
	\includegraphics[width=\columnwidth]{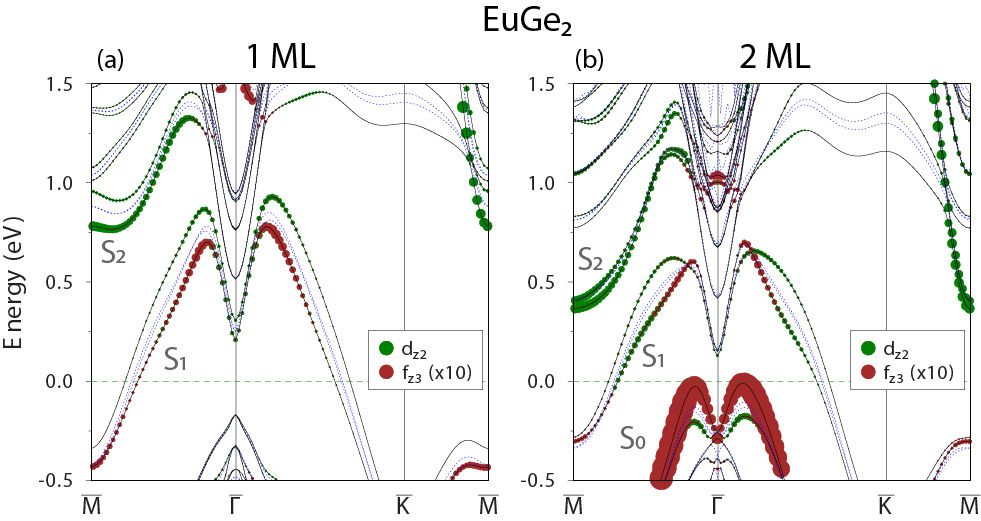}
	\caption{The effect of magnetism on the electronic band structure of 1 ML (a) and 2 ML (b) films of EuGe$_2$ metalloxene. The blue dashed and black solid curves simulate the spectra in non-magnetic and magnetic phases, respectively.}  
 \label{Fig-EuGe}
\end{figure}
%=======================================================

On the basis of DFT we define the preferred magnetization direction of 1 ML and 2 ML EuGe$_2$ films by calculating the total energies for the various magnetic configurations. As found the in-plane FM (FM$_\|$) and in-plane AFM (AFM$_\|$) are the most energetically favorable magnetic configurations for the 1 ML and 2 ML films, respectively, which is in agreement with experiment \cite{Sokolov2020MH}. For Eu 4$f$ orbitals the $U$ = 7.4 eV and $J$ = 1.1 eV parameters were used (which were adopted from previous work \cite{JPCL-EuS-Eremeev-2021}). Figure~\ref{Fig-EuGe}(a) shows the magnetic band structure of the 1 ML EuGe$_2$ film, demonstrating an emergence of noticeable exchange splitting in both the S1 and S2 surface bands, lifting the degeneracy at the time-reversal invariant momenta points ($\bar\Gamma$ and $\bar{\rm M}$). Orbital projections show that the exchange splitting of the S1, S2 and other bands comes from Gd $f-d$ hybridization, mediated mainly through the out-of-plane $f_{z^3}$ and $d_{z^2}$ components (shown as brown and green balls). As regards the 2 ML EuGe$_2$ film, the amplitude of exchange splitting here is mostly the same for most unoccupied bands, however, the unoccupied S0 band acquires larger $f_{z^3}$ orbital weights, and thus experiences increasing in the band splitting.

\end{document}